\documentstyle[aps,preprint,prb]{revtex}
\begin{document}
\tightenlines
\def\vk{\vec k} 
\def\vr{\vec r} 
\def\vq{\vec q} 
\def\br{{\bf r}}
\title{\bf Effect of Magnetic Impurities in Crystalline and Amorphous States of Superconductors}
\author{Mi-Ae Park }
\address{Department of Physics,  University of Puerto Rico at Humacao,\\
 Humacao, PR 00791}
\author{Kerim Savran and Yong-Jihn Kim}
\address{Department of Physics,  Bilkent University,\\
 06533 Bilkent, Ankara, Turkey}
\maketitle
\begin{abstract}

It has been observed that the effect of magnetic impurities in a superconductor
is drastically different depending on whether the host superconductor is in 
a crystalline or an amorphous state. 
Based on the recent theory of Kim and Overhauser,
it is shown that as the system is getting disordered, the initial slope of 
the $T_{c}$ depression is decreasing by a factor $\sqrt{\ell/\xi_{0}}$, 
when the mean free path $\ell$ becomes smaller than the BCS coherence length
$\xi_{0}$, which is in agreement with experimental findings. 
In addition, for a superconductor in a pure crystalline state the 
superconducting transition temperature $T_{c}$ drops sharply 
from about 50$\%$ of $T_{c0}$ (for a pure system) to zero 
near the critical impurity concentration. This {\sl pure limit behavior}
was indeed found by Roden and Zimmermeyer in crystalline Cd. 

\end{abstract}
\vskip 5pc
PACS numbers: 74.60.Mj, 74.20.-z, 74.90.+n

\vspace{1pc}

\noindent

\vfill\eject
\section{\bf Introduction} 

It has been well appreciated among experimentalists that the effect of magnetic 
imputities in superconductors depends strongly on whether the host
superconductor is in the crystalline state or in the amorphous state.$^{1-10}$
For instance, the initial slope of the
$T_{c}$ decrease due to magnetic impurities is not universal but dependent
on the sample quality and sample preparation methods.
This was not well understood.
Kim and Overhauser (KO)$^{11}$ have recently proposed a theory of magnetic 
impurity effect on superconductors, which anticipates the above experimental 
observations. To summarize, they obtained the following results:

(1) The initial slope of the $T_{c}$ decrease due to magnetic impurities depends
on the superconductor and therefore is not a universal constant.

(2) The reduction of $T_{c}$ by magnetic impurities is significantly
lessened whenever the mean free path $\ell$ becomes smaller than the BCS 
coherence length $\xi_{0}$.

(3) If the host superconductor is pure enough for exchange scattering
to dominate, $T_{c}$ drops suddenly from about 50 $\%$ of $T_{c0}$ (for the
pure metal) to zero near the critical impurity concentration. This 
may be called the {\sl pure limit behavior}.$^{7}$

The second result, called compensation phenomenon, 
has been observed by adding non-magnetic impurities$^{1,2}$ and radiation 
damage.$^{4,5,8}$ Indeed, Park, Lee, and Kim$^{12}$ showed that KO theory 
leads to a good fitting to the experimental data obtained from radiation 
damage.$^{8}$ There is also compelling evidence for the first result.
For example, the observed initial decrease of $T_{c}$ for superconductors 
as a function of the concentration c of the magnetic ions is bigger
in the crystalline state than that in the amorphous state
of superconductors. This behavior is also related to the second result.
In Table I, most of the values from the literature for the initial decrease 
of $T_{c}$ for the Zn-Mn system are listed, 
which shows this behavior spectacularly. Data are from Falke et al.$^{13}$
As can be seen, the initial slope of 
the $T_{c}$ decrease by magnetic impurities is not universal but dependent
on the sample quality and sample preparation methods.
Note that Zn-Mn alloys show all the Kondo anomalies at low temperatures.$^{19}$
The {\sl pure limit behavior} is hard to observe experimentally due to 
the metallurgical problems related to a very small solubility of 
magnetic impurities in non-transition metals. Also adding many magnetic
impurities may make the host superconductors disordered. Therefore, 
it is really remarkable that Roden and Zimmermeyer$^{6}$ confirmed the third 
result,  the {\sl pure limit behavior}, in crystalline Cadmium doped with 
dilute Mn atoms by quench condensation.
It is well known that sample preparation of thin-film alloys by evaporation
on cold substrates (quench condensation) is suitable for producing
alloys between otherwise insoluble components. They prepared 
(micro)crystalline and amorphous cadmium with dilute Mn impurities.
Remarkably, they found that a quench-condensed film of cadmium in the 
microcrystalline state shows an abrupt decrease of the transition temperature 
near the critical impurity concentration.  
Similar {\sl pure limit behavior} has been found in superfluid He-3 in 
aerogel.$^{20-22}$
In this case, aerogel acts like impurities and decreases the transition
temperature of the superfluid He-3.
It is straightforward to apply KO theory to p-wave
pairing states. More details will be published elsewhere.

In this paper we use KO theory to explain the difference of the
magnetic impurity effect in crystalline and amorphous states of
superconductors. We find a good agreement with the existing experimental
data. A brief review of experimental works is given in Sec. II, while KO
theory is described in Sec. III. Comparison with various experimental
data is given in Sec. IV. The {\sl pure limit behavior} will be emphasized.

\section{\bf Magnetic Impurity Effect in Crystalline and Amorphous States of Superconductors} 

In this Section, the experimental data for the effect of magnetic
impurities in a crystalline and an amorphous state of superconductors
are briefly reviewed.
Although there are already a few review articles on magnetic impurity
effect in superconductors,$^{23,24}$ this topic was not spotlighted before,
simply because the experimental data were not understood.
Nevertheless, it was observed by many experimentalists that 
the magnetic impurity effects are different for crystalline and amorphous 
states of superconductors. 
To illustrate, the initial decrease of $T_{c}$ for some superconductors 
as a function of the concentration c of the magnetic ions is summarized 
in Table II. The Table is from Buckel,$^{10}$ Wassermann,$^{3}$ and 
Schwidtal.$^{25}$
It is clear that the initial $T_{c}$ decrease depends on the sample
quality and is not the universal constant suggested by Abrikosov and
Gor'kov.$^{26}$ 
Note that In-Mn,$^{4,5,8}$ Sn-Mn,$^{28}$ Zn-Mn,$^{13}$ and Cd-Mn$^{35}$ show 
the Kondo anomalies at low temperatures.

Merriam, Liu, and Seraphim$^{1}$ were the first who found the difference.
They investigated the effect of dissolved Mn on superconductivity of
pure and impure In. They observed that the addition of a third element,
Pb or Sn, progressively decreases the effect of Mn and eliminates the effect
completely when the mean free path is decreased sufficiently enough.   
In other words, the $T_{c}$ depression arising from a paramagnetic
solute turned out to be mean-free-path dependent.
Boato, Gallinaro, and Rizzuto$^{2}$ confirmed the result.
It was also found that $T_{c}$ depression by transition
metal impurities in bulk metals and thin films leads very often to
different results.$^{3}$ For instance, broad scattering of the 
experimental $-dT_{c}/dc$ values was frequently obtained, presumably due to the
differences in the degree of disorder.
A review was given by Wassermann.$^{3}$  
On the other hand, Falke et al.$^{13}$ investigated transition temperature
depression in quench condensed Zn-Mn dilute alloy films
and compared it with bulk data. Their work gives good support 
to the equivalence of thin films and bulk material.  
To put it another way, even though the initial $T_{c}$ depression caused by 
magnetic impurities may be different for thin films and bulk material, 
a magnetic impurity may possess a stable 
magnetic moment whether it is in thin films or in bulk material.
Bauriedl and Heim$^{4}$ noted that the reason for the different
behavior of a magnetic impurity in crystalline and disordered materials
is lattice disorder.
The authors considered annealed In films implanted with 150 keV-Mn ions at
low temperatures and increased the lattice disorder by pre-implantation
of In ions, which led to the variation of the initial $T_{c}$-depression 
between 26 K/at $\%$ for the crystalline sample and 10 K/at $\%$ for the
heavily disordered sample.  
Hitzfeld and Heim$^{5}$ reported that the magnetic state of Mn in 
ion implanted In-Mn alloys is not so much affected by incorporating
oxygen (lattice disorder) but the superconducting properties changes
significantly, in agreement with Falke et al.$^{13}$: -d$T_{c}$/dc is changed 
from 24 to 18 K/at$\%$ if oxygen is added. 
Schlabitz and Zaplinski$^{7}$ reported on the influence of lattice
defects on the $T_{c}$-depression in dilute Zn-Mn single crystals.
Their measurements also show a much higher depression of $T_{c}$ 
for single crystals than for cold-rolled crystals and quench-condensed
films.
Hofmann, Bauriedl, and Ziemann$^{8}$ also observed compensation of the
paramagnetic impurity effect as a consequence of radiation damage.
Well annealed In-films implanted at low temperatures with Mn ions
lead to an initial slope of 50 K/at $\%$, whereas In-films irradiated
with high fluences of Ar ions before the Mn-implantation lead to a slope
of 39 K/at $\%$. In addition, 90$\%$ of the 2.2 K decrease in $T_{c}$
caused by Mn-implantation was suppressed by an Ar fluence of 2.2$\times$
$10^{16} cm^{-2}$.
Habisreuther et al.$^{9}$ reported on  an {\sl in situ} low-temperature 
ion-implantation study of Mn in crystalline $\beta$-Ga and amorphous 
$a$-Ga films. They found linear $T_{c}$ decreases in $a-$Ga films
with a slope of 3.4 K/at $\%$ and in $\beta-$Ga films with a slope 
of 7.0 K/at $\%$, (i.e., twice as large as in $a-$Ga).

Furthermore, Roden and Zimmermeyer$^{6}$ considered crystalline and 
amorphous cadmium with dilute Mn atoms. In the first case the initial 
depression of $T_{c}$ is $-dT_{c}/dc$=5.4 K/at $\%$ and in the second case
$-dT_{c}/dc$=2.65 K/at $\%$ in accordance with other results. 
Surprisingly, a sudden drop of $T_{c}$ in crystalline cadmium
near the critical concentration was observed. About 50 $\%$ of $T_{c0}$ 
was decreased to 
zero by adding additional tiny amounts of Mn atoms in the (micro)crystalline 
state, which has been predicted by Kim and Overhauser.
Since the transition temperature of  pure Cd in the crystalline state is 
0.9 K $(T_{c0})$, the critical Mn impurity concentration is so low 
($\sim$ 0.075 at $\%$) that the crystalline state is not much disturbed
by Mn atoms. Consequently, the {\sl pure limit behavior} of magnetic impurity
effect was observable.  
Zimmermeyer and Roden$^{36}$ also found similar behavior in microcrystalline
films of lead doped with Mn, but with a peak just before $T_{c}$ drops to 
zero suddenly. 
The critical concentration is $\sim$ 0.4 at $\%$.  
In this case, since the initial $T_{c}$ depression is not linear as a 
function of Mn concentration, there seems to be some solubility problem.

\vspace{1pc}

\section{\bf Theory of Kim and Overhauser} 

\subsection{\bf Ground state wavefunction} 

For a homogeneous system, the BCS wavefunction is given by$^{37,38}$
\begin{equation}
\tilde{\phi}=\prod_{k}(u_{k}+v_{k}a_{k\uparrow}^{\dagger}
a_{-k\downarrow}^{\dagger})\phi_{0}
\end{equation} 
where the operator $a_{k\alpha}^{\dagger}$ creates an electron in the state
$(k\alpha)$ (with the energy $\epsilon_{k}$) when operating on the vacuum state designated by $\phi_{0}$.
Note that $\tilde{\phi}$ is an approximation of $\phi_{N}$,
\begin{equation}
\phi_{N}=A[\phi(r_{1}-r_{2})\cdots
\phi(r_{N-1}-r_{N})(1\uparrow)(2\downarrow)\cdots(N-1\uparrow)(N\downarrow)]
\end{equation} 
where
\begin{equation}
\phi(r)=\sum_{k}{v_{k}\over u_{k}}e^{i{\bf k}\cdot{\bf r}}
\end{equation} 
and both wavefunctions lead to the same result for a large system.
Nevertheless, $\phi_{N}$ is more helpful for understanding the underlying
physics related to the magnetic impurity effect in superconductors:
we are concerned with a bound state of Cooper pairs in a BCS condensate.
It should be noticed that the (bounded) pair wavefunction $\phi(r)$ and 
the BCS pair-correlation amplitude $I(r)^{37}$ are basically the same 
for large N:
\begin{equation}
\phi(r)=\sum_{k}{\Delta_{k}\over \epsilon_{k}+E_{k}}e^{i{\bf k}\cdot{\bf r}},
\end{equation} 
\begin{equation}
I(r)=\sum_{k}{\Delta_{k}\over 2E_{k}}e^{i{\bf k}\cdot{\bf r}}\sim \Delta K_{0}({r\over \pi\xi_{0}}),
\end{equation} 
where
\begin{equation}
E_{k}=\sqrt{\epsilon_{k}^{2}+\Delta_{k}^{2}}.
\end{equation} 
Here $K_{0}$ is a modified Bessel function which decays rapidly when $r>\pi\xi_{0}$.

In the presence of magnetic impurities, 
BCS pairing must employ degenerate
partners which have the exchange scattering (due to magnetic impurities) 
built in
because the strength of exchange scattering $J$ is much larger than
the binding energy. This scattered state representation was first introduced
by Anderson$^{39}$ in his theory of dirty superconductors.
Accordingly, the corresponding wavefunctions are
\begin{equation}
\tilde{\phi'}=\prod_{n}(u_{n}+v_{n}a_{n\uparrow}^{\dagger}
a_{\bar{n}\downarrow}^{\dagger})\phi_{0}
\end{equation} 
and
\begin{equation}
\phi'_{N}=A[\phi'(r_{1},r_{2}) \phi'(r_{3},r_{4})\cdots
\phi'(r_{N-1},r_{N})]
\end{equation} 
where
\begin{equation}
\phi'(r_{1},r_{2})=\sum_{n}{v_{n}\over u_{n}}\psi_{n\uparrow}(r_{1})
\psi_{{\bar n}\downarrow}(r_{2}).
\end{equation} 
Here $\psi_{n\uparrow}$ and $\psi_{{\bar n}\downarrow}$ denote the exact eigenstate and its
degenerate partner, respectively. 
It is clear from the pair wavefunction $\phi'(r_{1},r_{2})$ that only 
the magnetic impurities within $\xi_{0}$ of a Cooper pair's 
center of mass can diminish the pairing interaction.

\subsection{\bf Phonon-mediated matrix element} 

Now we need to determine the scattered state
$\psi_{n}$ and the phonon-mediated matrix element $V_{nn'}$.
The magnetic interaction between a conduction electron at $\bf r$ and a 
magnetic impurity (having spin $\bf S$), located at ${\bf R}_{i}$, is given by
\begin{equation}
H_{m}({\bf r})=J{\bf s}\cdot{\bf S}_{i}v_{o}\delta({\bf r}-{\bf R}_{i}),
\end{equation} 
where ${\bf s}={1\over 2}\sigma$ and $v_{o}$ is the atomic volume.
The scattered basis state which carries the 
label, $n\alpha={\vec k}\alpha$, is then
\begin{equation}
\psi_{n\alpha=\vk\alpha} = N_{\vk} [ e^{i\vk\cdot \vr}\alpha
+ \sum_{\vq}e^{i(\vk + \vq)\cdot \vr}(W_{\vk\vq}\beta + W_{\vk\vq}^{'}
\alpha)], 
\end{equation}
where,  
\begin{equation}
W_{\vk\vq} = {{1 \over 2}J\overline{S}v_{o} \over 
\epsilon_{\vk} - \epsilon_{\vk+\vq}}\sum_{j}sin \chi_{j} e^{i\phi_{j}
-i\vq\cdot {\bf R}_{j} }
\end{equation}
and,
\begin{equation}
W_{\vk\vq}' = {{1 \over 2}J\overline{S}v_{o} \over
\epsilon_{\vk} - \epsilon_{\vk + \vq}}\sum_{j} cos \chi_{j} e^{-i\vq\cdot
{\bf R}_{j}}. 
\end{equation}
$\chi_{j}$ and $\phi_{j}$ are the polar and azimuthal angles of the spin
${\bf S}_{j}$ at ${\bf R}_{j}$ and $\overline{S}=\sqrt{S(S+1)}$. 
The perturbed basis state for the degenerate partner of (11)
is:
\begin{equation}
\psi_{{\bar n}\beta=-\vk\beta} = N_{\vk}[e^{-i\vk\cdot\vr}\beta
+ \sum_{\vq}e^{-i(\vk + \vq)\cdot \vr} (W_{\vk\vq}^{*}\alpha - W_{\vk\vq}
^{'*}\beta)]. 
\end{equation}

At each point $\vr$, the two spins of the degenerate partner become
canted by the mixing of the plane wave and spherical-wavelet
component. Consequently, the BCS condensate is forced to have a triplet
component because of the canting caused by the exchange scattering.
The phonon-mediated matrix element between the canted basis pairs is (to order $J^{2}$)
\begin{equation}
V_{nn'}\equiv  V_{\vk'\vk} = - V < cos\theta_{\vk'}(\vr)> < cos\theta_{\vk}(\vr)>,
\end{equation}
where $\theta$ is the canting angle.
The angular brackets indicate both a spatial and impurity average.
It is then given
\begin{equation}
<cos\theta_{\vk}(\vr)>\ \cong \ 1 -2|W_{\vk}|^{2}, 
\end{equation}
where $|W_{\vk}|^{2}$ is the relative probability contained in the
virtual spherical waves surrounding the magnetic solutes (compared
to the plane-wave part).
From Eqs. (11)-(13) we obtain
\begin{equation}
|W_{\vk}|^{2} = {J^{2}m^{2}{\bar S}^{2}c_{m}R\over 8\pi n\hbar^{4}},
\end{equation}
where $c_{m}$ is the magnetic solute fraction.
Because the pair-correlation amplitude 
falls exponentially as $exp(-r/\pi\xi_{0})$$^{37}$ at $T=0$ and
as $exp(-r/3.5\xi_{0})$$^{40}$ near $T_{c}$, 
we set
\begin{equation}
R = {3.5\over 2}\xi_{0}. 
\end{equation}
Then one finds
\begin{equation}
<cos\theta> = 1 - {3.5\xi_{0} \over 2 \ell_{s}}, 
\end{equation}
where $\ell_{s} = v_{F}\tau_{s}$ is the mean free path for exchange 
scattering only.

\subsection{\bf BCS $T_{c}$ equation} 

The resulting BCS gap equation, near $T_{c}$, is given by
\begin{equation}
\overline{\Delta_{\vk}}= -\sum_{\vk'} \overline{V_{\vk,\vk'}}
{\overline{\Delta_{\vk'}}\over 2\overline{\epsilon_{\vk'}}}tanh({\overline{\epsilon_{\vk'}}\over 2T}).
\end{equation}
Here $\overline{\Delta_{\vk}}$ is 
the impurity averaged value of the gap parameter 
whereas $\overline{\epsilon_{\vk}}$ is that of
the electron energy.
The BCS $T_{c}$ equation still applies after a modification of the effective 
coupling constant according to Eq. (15):
\begin{equation}
\lambda_{eff} = \lambda <cos\theta>^{2}, 
\end{equation}
where $\lambda $ is $N_{o}V.$ Accordingly, the BCS $T_{c}$ equation
is now,
\begin{equation}
k_{B}T_{c} = 1.13\hbar\omega_{D}e^{-{1\over \lambda_{eff}}}. 
\end{equation}
The initial slope is given
\begin{equation}
k_{B}(\Delta T_{c}) \cong -{0.63\hbar \over \lambda\tau_{s}}. 
\end{equation}
The factor $1/\lambda$ shows that the initial slope depends on the
superconductor and is not  a universal constant.
For an extended range of solute concentration,
KO find 
\begin{equation}
<cos\theta> = {1\over 2} + {1\over 2}[1 + 5({u\over 2})^{2}]^{-1}
e^{-2u}, 
\end{equation}
where
\begin{equation}
u \equiv 3.5\xi_{eff}/2\ell_{s}. 
\end{equation}
We have replaced $\xi_{0}$ by the effective coherence length  $\xi_{eff}$
which is explained below.

\subsection{\bf Change of the initial slope of the $T_{c}$ decrease} 

When the conduction electrons have a mean free path $\ell$ which is smaller
than the coherence length $\xi_{0}$ (for a pure superconductor), the
effective coherence length is 
\begin{equation}
\xi_{eff} \approx \sqrt{\ell\xi_{0}}. 
\end{equation}
For a superconductor which has ordinary impurities as well
as magnetic impurities, the total mean-free path $\ell$ is given by 
\begin{equation}
{1\over \ell} = {1\over \ell_{s}} + {1\over \ell_{0}}, 
\end{equation}
where $\ell_{0}$ is the mean free path for the potential scattering.
It is evident from Eq. (26) that the potential scattering 
profoundly affects the paramagnetic impurity effect. 
Consequently, the initial slope of the $T_{c}$ depression is decreased in
the following way:
\begin{equation}
k_{B}(\Delta T_{c}) \cong -{0.63\hbar \over \lambda\tau_{s}}\sqrt{\ell\over \xi_{0}}. 
\end{equation}
This explains the broad scattering of the experimental $-dT_{c}/dc$ values.
In other words, the size of the Cooper pair is reduced by the potential
scattering and the reduced Cooper pair sees a smaller number of
magnetic impurities. Accordingly, the magnetic impurity effect is
partially suppressed, leading to the decrease of the initial slope of 
the $T_{c}$ depression.  

Figure 1 shows the different behavior of the $T_{c}$ depression due to magnetic 
impurities in the pure crystalline state and in the amorphous or disordered 
state of superconductors.
We used $T_{co}=1.0K$, $v_{F}=1.5\times 10^{8}cm/sec$, and $\omega_{D}=250K$.
We also assumed the relation between $\ell_{s}$ and magnetic impurity
concentration c: $\ell_{s}=10^{5}/c (\AA)$. 
Here $c$ is measured in at $\%$.  Since the exchange scattering
cross-section is usually 20-200 times smaller than that for the  potential 
scattering,$^{11}$ this assumption seems to be reasonable. 
For the pure crystalline state $T_{c}$ drops to zero suddenly when $T_{c}$
is decreased to about 50 \% of $T_{c0}$ of the pure system, which may be
called {\sl pure limit behavior}.
As the mean free path $\ell$ is decreased due to disorder,
the initial $T_{c}$ depression is weakened and $T_{c}$ drops to zero
more slowly near the critical concentration.  

\vspace{1pc}

\section{\bf Comparison with Experiment} 

The overall agreement between KO theory and the existing experimental
data is impressive. We focus on the experiments which 
investigated the
difference of the magnetic impurity effect in pure crystalline
state and amorphous or disordered state of superconductors.

\subsection{\bf {\sl Pure limit behavior}: Roden and Zimmermeyer's Experiment} 

Roden and Zimmermeyer$^{6}$ prepared alloys of Cd with dilute Mn impurities
by quench condensation. Quench condensation
produces a variety of states of the alloy: in particular,
one can get a microcrystalline and an amorphous state.
A quench-condensed film of Cd in the microcrystalline state shows a 
higher $T_{c}$ ($=0.9K$) than the bulk material ($T_{c}=0.55K$) and 
a further increase of
$T_{c}$ ($=1.15K$) is obtained in the amorphous state.  
Amorphous Cd film was obtained by adding Cu atoms.
Like other nontransition metals deposited in an ordinary
high-vacuum system, quench-condensed Cd film is crystalline with small 
crystallites.$^{41}$

Now we compare KO theory with Roden and Zimmermeyer's experiment.
Figure 2 shows $T_{c}$ versus magnetic impurity concentration c in the
microcrystalline CdMn.
The solid line is theoretical curve obtained from Eq. (22).
The transition temperature $T_{c0}$ of pure Cd in this state is 0.9K.
While the initial depression of $T_{c}$ is linear in c with a value
of $-dT_{c}/dc =5.4K/at\%$, above 0.05$\%$ the depression becomes much
more stronger than linear, which agrees with KO theory. Arrows denote
that no superconductivity was found up to 70mK. 
For theoretical fitting we used $T_{c0}=0.904K$, $\omega_{D}=209K$,
and $v_{F}=1.62\times 10^{8}cm/sec$.$^{42}$
We emphasize that there is no free parameter. 
But, in the absence of experimental data we assumed 
$\ell_{s}=9\times 10^{5}/c\ (\AA)$.
As can be seen, the agreement between the experimental data and the
theoretical curve is very good.

Figure 3 shows $T_{c}$ vs. c for the amorphous CdCuMn.
The solid line was obtained from Eqs. (22) and (26).
The decrease of $T_{c}$ for smaller c is again linear but with a
much lower $-dT_{c}/dc=2.65 K/at\%$. In the amorphous state $T_{c0}$ is
about 1.18K. Since the residual resistivity data are not available,
we assumed that the mean free path for the potential scattering  
is $\ell_{0}=4500\AA$ which is reasonable. We used the same values for 
$\omega_{D}$ and $v_{F}$ as in Fig. 2.
Again we find a good fitting to experimental data.

\subsection{\bf Change of the initial slope of the $T_{c}$ depression} 

Schlabitz and Zaplinski$^{7}$ reported measurements of the $T_{c}$-depression
of ZnMn single crystals. In particular, they investigated the influence of 
lattice defects on the $T_{c}$-depression in dilute ZnMn single
crystals. They demonstrated linear behavior up to a concentration
of 10 ppm with a slope of 630 K/at$\%$. This value is twice that of
other measurements. As a result, they suggested that the $T_{c}$-depression
can be enhanced strongly by eliminating the lattice defects.

Figure 4 shows the reduced transition temperature, $T_{c}/T_{c0}$, as 
a function of Mn concentration for ZnMn samples. The dashed lines, 
taken from the
other measurements,$^{13}$ give the $T_{c}$-depression of: a) quench-condensed
films, and b) cold rolled bulk material. The filled points represent
the $T_{c}$-values of the ZnMn single crystals. 
The filled squares are the data of quench condensed thin 
films, while the filled triangles are the data of quench condensed thin 
films after annealing at $300K$ for 14 hours. 
Since annealing leads to an increased order of the lattice,$^{41}$ 
it is clear that the initial slope of the $T_{c}$ decrease is decreasing
as the system is getting disordered.

The solid line is the theoretical curve obtained from the initial slope,
$-dT_{c}/dc=630K/at \%$ with $T_{c0}=0.9K$:
\begin{equation}
k_{B} T_{c} \cong k_{B} T_{c0} - {0.63\hbar \over \lambda\tau_{s}}.
\end{equation}
This expression agrees very well with the exact BCS $T_{c}$ equation, Eq. (22),
up to 25 $\%$ of the critical impurity concentration. 
The dashed lines (a) and (b)
can also be reproduced from the theoretical formula, Eq. (28), 
\begin{equation}
k_{B} T_{c} \cong k_{B} T_{c0} - {0.63\hbar \over \lambda\tau_{s}}\sqrt{\ell\over \xi_{0}},
\end{equation}
for the 
initial $T_{c}$ depression in the disordered state of superconductors with 
$(a)\ \ell=7520\AA, T_{c0}=0.83K$,$^{13}$ and 
$(b)\ \ell=3390\AA, T_{c0}=1.51K$,$^{13}$ respectively.  
Here $T_{c0}$ values are the experimental results.$^{13}$
Therefore, the change of the initial slope of the $T_{c}$ decrease may be
explained in terms of the change of the Cooper pair size 
caused by the variation of the  mean free path $\ell$.  
We used $\omega_{D}=327K$, and $v_{F}=1.82\times 10^{8} cm/sec$.$^{42}$ 
The sudden drop of $T_{c}$ near the critical concentration
is not pronounced though, presumably because of the smallness of the 
critical concentration. Since there are not many magnetic imputities in the
Zn matrix, the distribution of Mn may be atomically disperse but
macroscopically inhomogeneous.
Then, the {\sl pure limit behavior} may not be observable. 

Bauriedl and Heim$^{5}$ investigated the influence of lattice disorder on
the magnetic properties of InMn alloys. Crystalline In films
were implanted by Mn ions. The amount of lattice disorder
was changed in a very controlled way by pre-implantation of
indium with its own ions, which was very effective in 
producing disordered films. 

Figure 5 shows the transition temperatures for InMn alloys with
increasing lattice disorder from 1 to 3 by pre-implantation of 
$In^{+}$ ions: 1 0ppm; 2 2660ppm; 3 18,710ppm.
These ions have an intensive damaging effect, resulting in an increased
residual resistivity and an enhanced transition temperature $T_{c0}$.$^{5}$
Notice that the initial slope decreases as the system is more
disordered.
The solid lines are the theoretical results from Eq. (28) with
$1\ \ell=1050\AA$, $2\ \ell=700\AA$ and $3\ \ell=150\AA$.
It is necessary to emphasize that the change of the initial slope due to
the enhanced $T_{c0}$ (Eq. (23)) is not enough to explain the experimental data.
We assumed the initial slope $-dT_{c}/dc=53K/at \%$ for a pure system.$^{31}$ 
We also used $\omega_{D}=108K$ and $v_{F}=1.74\times 10^{8}$.$^{42}$
We find good agreements between theory and experiment.

Finally, Habisreuther et al.$^{9}$ investigated the magnetic behavior of Mn
in crystalline $\beta-$Ga and amorphous $a-$Ga films. 
Mn ions were implanted at low temperature ($T<10K$). 
The amorphous $a-$Ga exhibits a rather high transition temperature
with typical values between 8.1 and 8.4 K, while the crystalline
$\beta-$Ga phase shows transition temperature of $T_{c}=6.3K$.

Figure 6 shows changes of the superconducting transition
temperature $\Delta T_{c}$ produced by Mn implantation into
amorphous $a-$Ga films and crystalline $\beta-$Ga films as a
function of the impurity concentrations.
Note that the initial slope 3.4 K/at $\%$ in amorphous $a-$Ga  is
about half of that (7.0 K/at $\%$) in crystalline $\beta-$Ga films.   
Theoretical curves represent the initial slope formulas, Eq. (23)
and (28) with $-dT_{c}/dc=7K/ at \%$, $\ell=\infty$ for $\beta-$Ga, and  
with $-dT_{c}/dc=7K/at \%$, $\ell=600\AA$ for $a-$Ga.
We used $\omega_{D}=320K$ and $v_{F}=1.91\times 10^{8} cm/sec$.$^{42}$
A good fitting to the experimental data is obtained.

\vspace{1pc}

\section{\bf Discussion} 

It is clear that a systematic experimental study of the effect of
magnetic impurities in crystalline and amorphous superconductors is
necessary. In particular, the {\sl pure limit behavior} in a crystalline 
state of superconductors and the change of the initial slope due to
disordering need more careful studies.
This study may shed a new light on the old question of whether a transition 
metal impurity possesses a stable local magnetic moment
within a metallic host.   

The observed {\sl pure limit behavior} in the superfluid He-3 in aerogel
may be compared with that in crystalline superconductors including Cd.
In superfluid He-3 aerogel does not disturb the liquid state
of He-3 significantly, whereas in superconductors adding magnetic impurities 
may damage the crystalline state of the superconductors, resulting in the
difficulty in observing the {\sl pure limit behavior}. 

In our theoretical fitting we guessed the mean free path $\ell$ because
the experimental residual resistivity data were not available. 
If the residual resistivity is given, the mean free 
path $\ell$ can be determined from the Drude formula. 
It is interesting to note that the initial $T_{c}$ depression also provides 
a way to estimate the mean free path $\ell$.

In this study, weak-coupling BCS theory is used to investigate the effect 
of magnetic impurities in superconductors. It is straightforward to
extend this study to the strong-coupling theory.$^{43,44}$
To do that, pairing of the degenerate scattered state 
partners is also needed.$^{45}$ The result will then basically be the 
same as that of the weak-coupling theory. More details will be 
published elsewhere.

\section{\bf Conclusion} 

The effect of magnetic impurities in crystalline and amorphous states of
superconductors has been studied theoretically. 
The {\sl pure limit behavior} in crystalline Cd observed by Roden and 
Zimmermeyer and the decrease of 
the initial slope of the $T_{c}$ depression due to disorder have been 
explained.
In particular, the initial slope of the $T_{c}$ decrease is decreasing
by a factor $\sqrt{\ell/\xi_{0}}$ as the system is getting disordered.
We suggest that a more systematic experimental investigation is necessary for 
the different behavior of the magnetic impurities in crystalline and amorphous
superconductors. Such an investigation may also be important for understanding 
whether a transition metal impurity possesses a magnetic moment in a metallic 
host or not.

\vspace{1pc}

\centerline{\bf ACKNOWLEDGMENTS}

Y.J.K. is grateful to Department of Physics at UPR-Humacao where this work
was finished. M. Park thanks the FOPI at the University of 
Puerto Rico-Humacao for release time.

\vfill\eject
{\bf Table I.} Values for the initial depression $-(dT_{c}/dc)_{initial}$ of the $T_{c}$ of Zn with different concentrations of Mn. Data are from Falke et al., Ref. 13.

\vspace{2pc}
\hspace{2pc}

\begin{tabular}{lrr} \hline \hline
{$-(dT_{c}/dc)_{initial}$ in [K/at \%]} \hspace{2pc}  & sample\hspace{2pc} & \hspace{2pc} Reference  \\ \hline

170 \hspace{1pc} & bulk \hspace{2pc} & \hspace{1pc} [14] (1964)\\ 

315 \hspace{1pc} & bulk \hspace{2pc} & \hspace{1pc} [2] (1966)\\ 

$>$300 \hspace{1pc} & bulk \hspace{2pc} & \hspace{1pc} [15] (1968)\\ 

260 (290) \hspace{1pc} & bulk \hspace{2pc} & \hspace{1pc} [16] (1971)\\ 

300 \hspace{1pc} & bulk \hspace{2pc} & \hspace{1pc} [17] (1972)\\ 

630 \hspace{1pc} & single crystal \hspace{2pc} & \hspace{1pc} [7] (1975)\\ 

215 \hspace{1pc} & thin film \hspace{2pc} & \hspace{1pc} [18] (1967)\\ 

285 \hspace{1pc} & thin film \hspace{2pc} & \hspace{1pc} [13] (1967)\\ 
\hline\hline
\end{tabular}

\vfill\eject

{\bf Table II.} Reduction in the $T_{c}$ of some superconductors by magnetic impurities. Data are from Buckel, Ref. 10, Wassermann, Ref. 3, and Schwidtal, Ref. 25.

\vspace{2pc}
\hspace{2pc}

\begin{tabular}{lrr} \hline \hline
{Superconductor}\hspace{2pc}  & Additive\hspace{2pc} & \hspace{2pc} {-$dT_{c}/dc$ in K/atom \%} \\ \hline

\hspace{1pc} Pb & Mn \hspace{2pc} & \hspace{1pc} 21$^{*}$ a),\hspace{1pc} 16$^{**}$ b) \\ 

 \hspace{1pc} Sn & Mn \hspace{2pc} & \hspace{1pc} 69$^{*}$ c),\hspace{1pc} 14$^{**}$ b)\\ 

 \hspace{1pc} Zn & Mn \hspace{2pc} & \hspace{1pc} 315 d),\hspace{1pc} 285$^{*}$ e),\hspace{1pc} 343$^{**}$ f),\hspace{1pc} 630 g) \\ 

\hspace{1pc} Zn & Cr \hspace{2pc} & \hspace{1pc} 170 d),\hspace{1pc} 90-200 h)\\ 
 \hspace{1pc} Cd & Mn \hspace{2pc} & \hspace{1pc} 44 i),\hspace{1pc} 5.4$^{*}$ j)\\ 

 \hspace{1pc} In & Mn \hspace{2pc} & \hspace{1pc} 25 k),\hspace{1pc} 53$^{*}$ l),\hspace{1pc} 50$^{**}$ m),\hspace{1pc} 100 n)\\ 

 \hspace{1pc} In & Fe \hspace{2pc} & \hspace{1pc} 2.5 l),\hspace{1pc} 2.0 o) \\ 

 \hspace{1pc} La & Gd \hspace{2pc} & \hspace{1pc} 5.1$^{*}$ p),\hspace{1pc} 4.5$^{**}$ q)\\ 
\hline\hline

\end{tabular}

$^{*}$ quench-condensed films

$^{**}$ ion implantation at low temperatures

References: a): [27]; b):[28]; c):[29]; d):[2]; e):[13]; f):[30]; g):[7]; 

h):[3]; i):[18]; j):[6]; k):[4]; l):[31]; m):[8]; n):[1]; o):[32]; p):[33]; q):[34];

\vfill\eject

\begin{figure}
\caption{Variation of $T_{c}$ with magnetic impurity concentration for pure and impure superconductors. $\ell_{o}$ denotes the mean free path for the potential scattering.} 
\end{figure}

\begin{figure}
\caption{Comparison of the experimental data for CdMn in the microcrystalline state with the KO theory. Data are from Roden and Zimmermeyer, Ref. 6.} 
\end{figure}

\begin{figure}
\caption{Comparison of the experimental data for CdMn in the amorphous state with the KO theory. Data are from Roden and Zimmermeyer, Ref. 6.} 
\end{figure}

\begin{figure}
\caption{Reduced transition temperature versus Mn concentration for ZnMn. 
The solid line represents the theoretical curve obtained from Eq. (29).
Line (a): Data of thin films from Ref. 13, line (b): Data of cold rolled bulk material from Refs 2 and 16. Data are from Schlabitz and Zaplinski, Ref. 7.} 
\end{figure}

\begin{figure}
\caption{Calculated transition temperatures for implanted InMn alloys. Increasing lattice disorder from 1 to 3 has been produced by pre-implantation of In ions: 1 0ppm, 2 2660 ppm, 3 18,710 ppm. Data are from Bauriedl and Heim, Ref. 4.} 
\end{figure}

\begin{figure}
\caption{Calculated changes of the superconducting transition temperature $\Delta T_{c}$ versus impurity concentration for Mn-implanted amorphous $a-$Ga and crystalline $\beta-$Ga. Data are from Habisreuther et al., Ref. 9.} 
\end{figure}

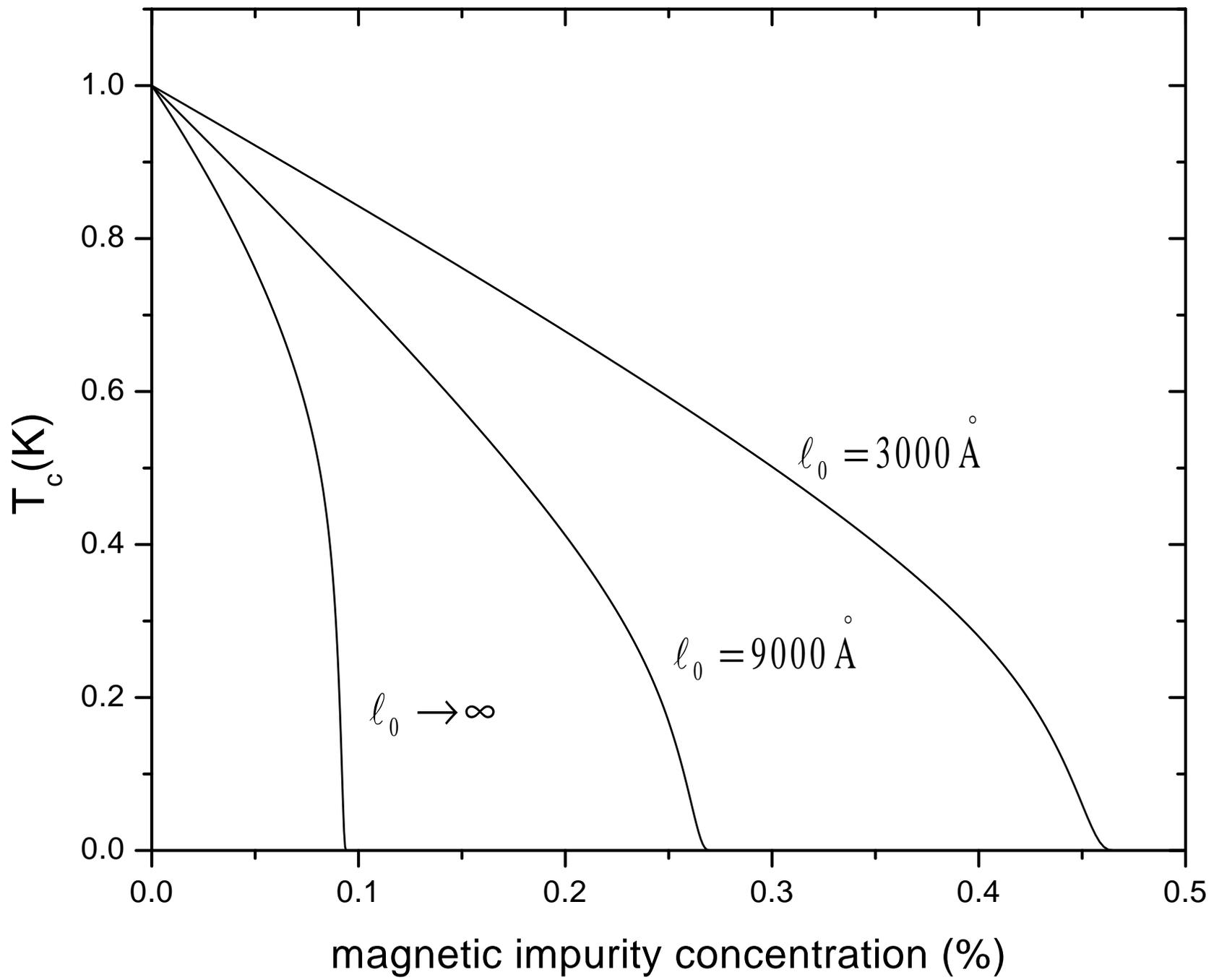
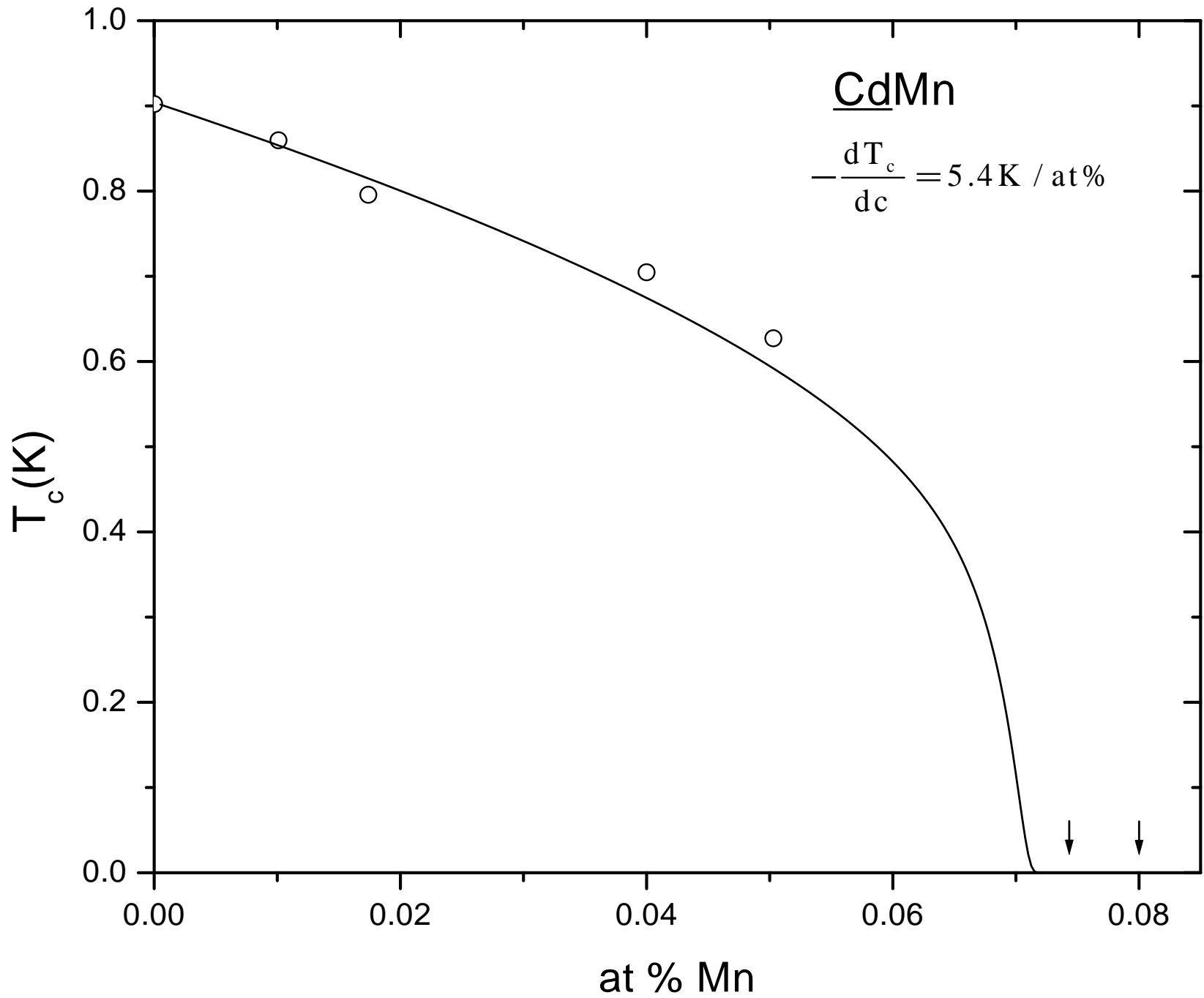
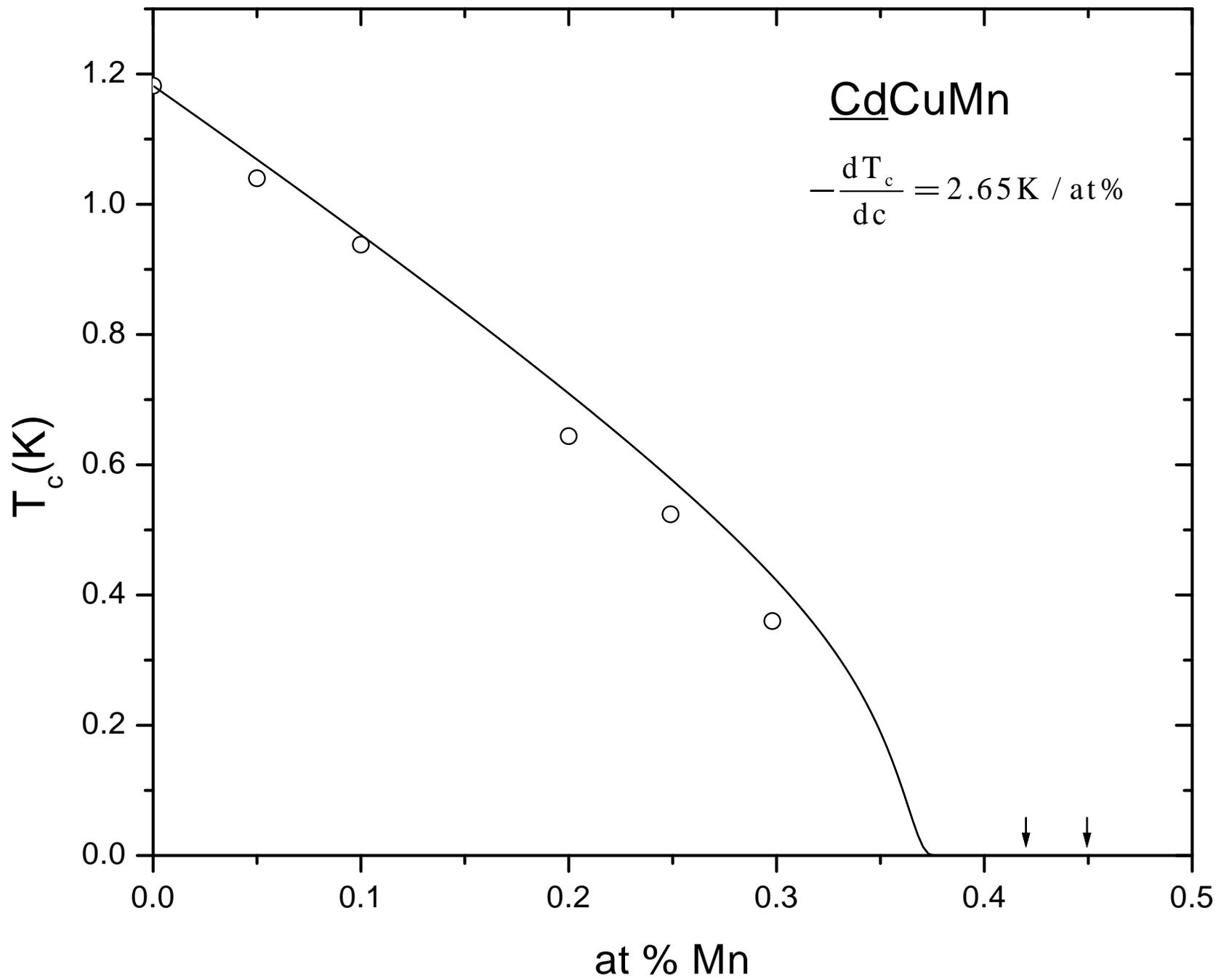
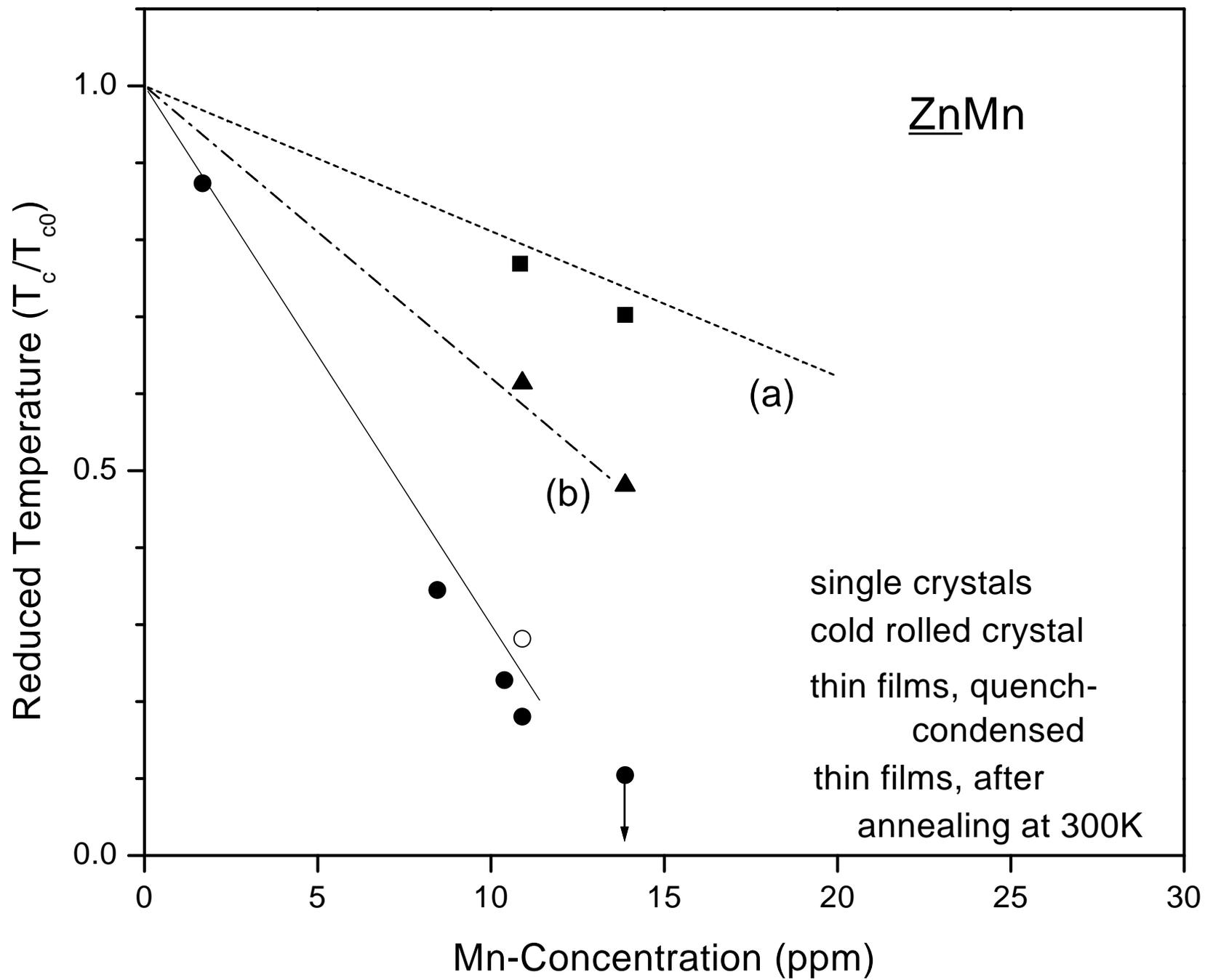
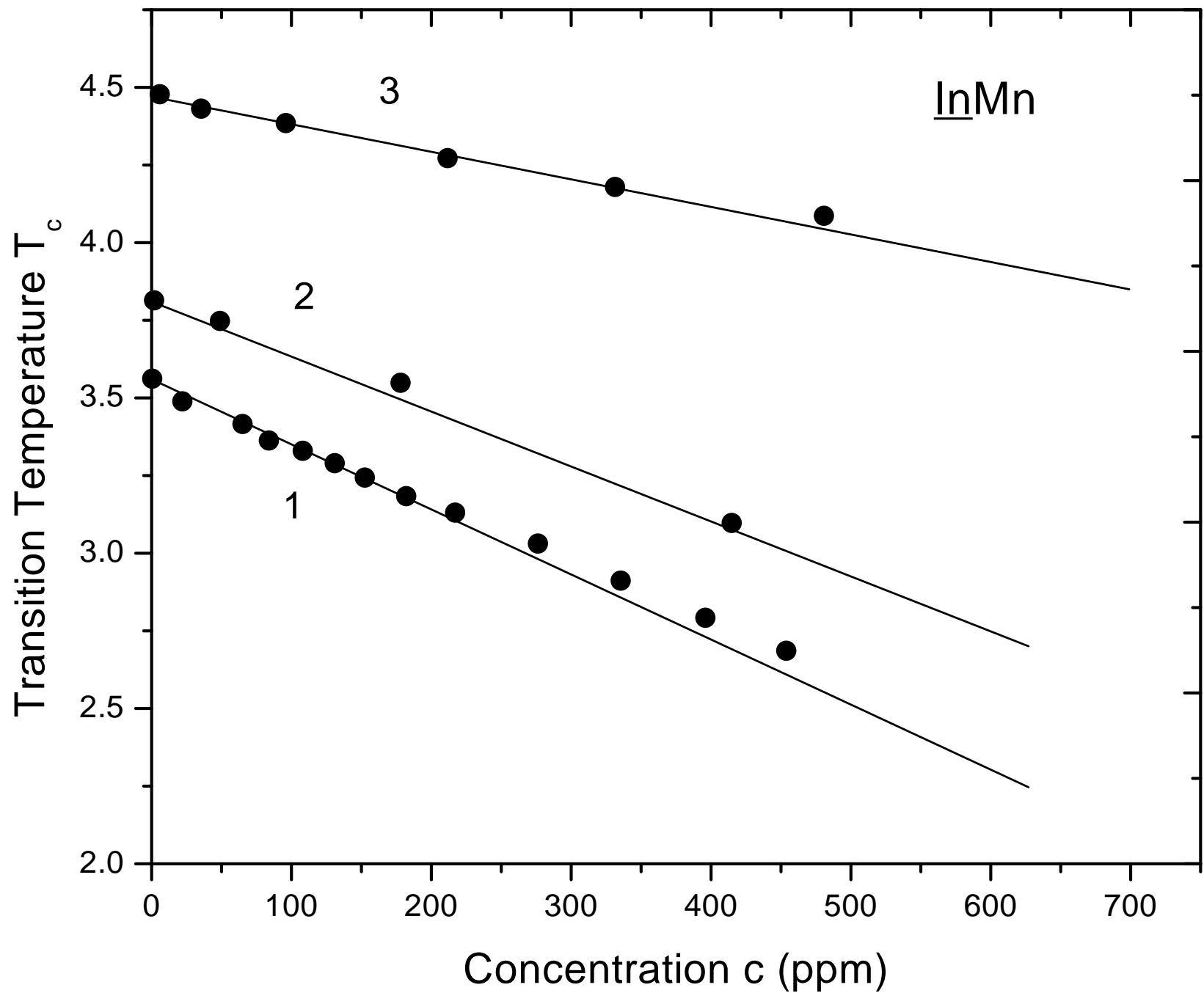
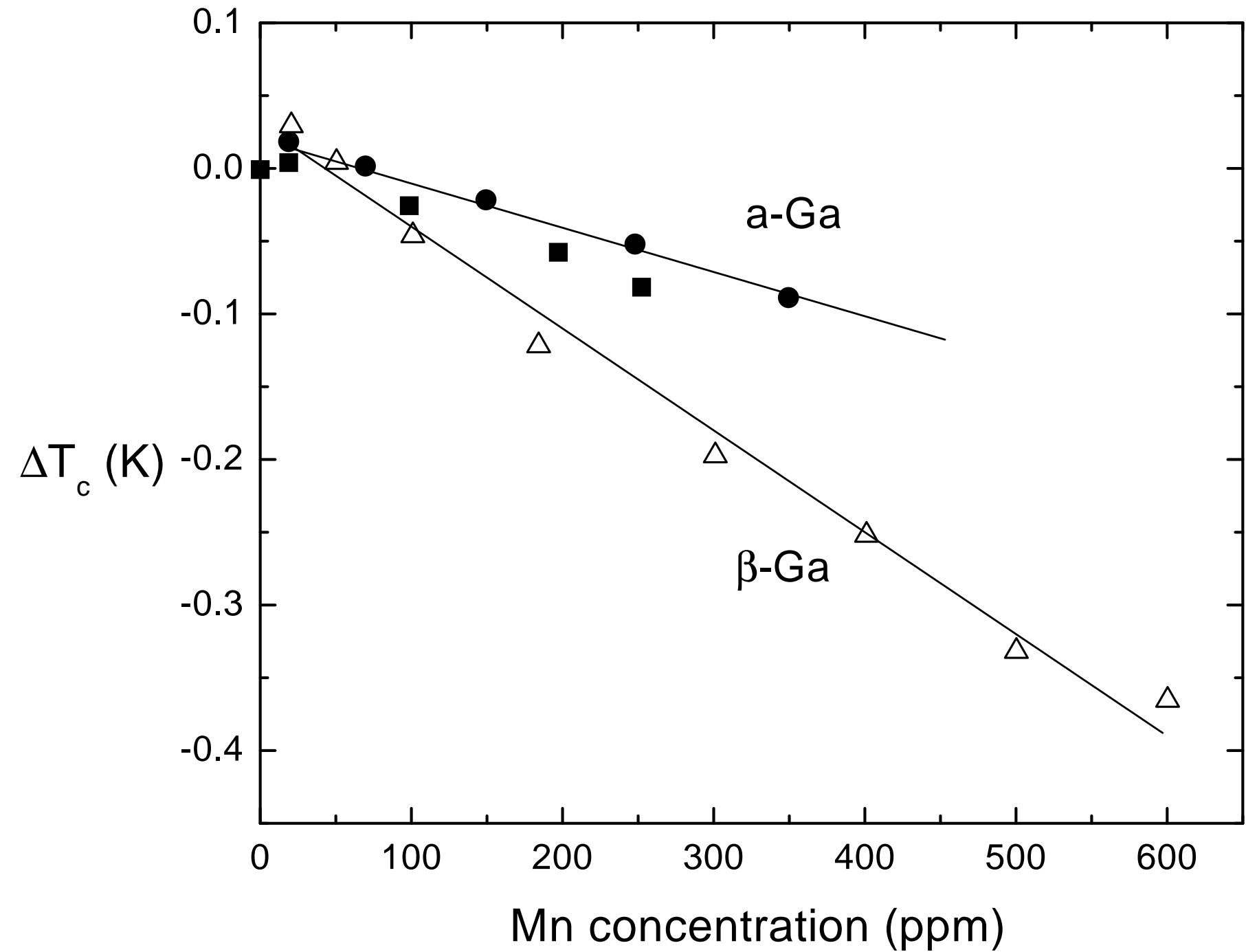
\end{document}